\documentclass[12pt]{article}

\usepackage{amsmath}
\usepackage{amsfonts}

\def\be{\begin{equation}}
\def\ee{\end{equation}}

\newtheorem{theorem}{Theorem}
\newtheorem{lemma}[theorem]{Lemma}

\newtheorem{definition}[theorem]{Definition}

\begin{document}

\title{Towards ultrametric theory of turbulence}

\author{ S.V.Kozyrev}

\maketitle

\centerline{\it Steklov Mathematical Institute}

\centerline{\it Gubkin St. 8, 119991 Moscow, Russia}

\begin{abstract}
Relation of ultrametric analysis, wavelet theory and cascade models
of turbulence is discussed. We construct the explicit solutions for
the nonlinear ultrametric integral equation with quadratic
nonlinearity. These solutions are built by means of the recurrent
hierarchical procedure which is analogous to the procedure used for
the cascade models of turbulence.
\end{abstract}

\section{Introduction}

$p$--Adic analysis and mathematical physics attract a lot of
attention, see \cite{VVZ}, \cite{Andr1}, \cite{Kochubei}.  $p$--Adic
pseudodifferential operators (in particular, the Vladimirov operator
$D^{\alpha}$) and pseudodifferential equations (analogues of the
equations of mathematical physics) were studied. $p$--Adic wavelets
were introduced and the relation of wavelet analysis to spectral
theory of $p$--adic pseudodifferential operators was described in
\cite{wavelets}. Analysis of pseudodifferential operators and
wavelets on general (locally compact) ultrametric spaces was
developed in \cite{Izv}, \cite{ACHA}, \cite{MathSbornik}.

In the present paper we construct families of explicit solutions for
the following nonlinear integro--differential equation:
$$
\frac{\partial}{\partial t}v(x,t)+\int\int F({\rm
sup}(x,\xi,\eta))v(\xi,t)\left(v(\eta,t)-v(x,t)\right)d\nu(\xi)
d\nu(\eta)+
$$
$$
+\int G({\rm sup}(x,\xi))(v(x,t)-v(\xi,t))d\nu(\xi)=0.
$$
Here the argument $x\in X$ belongs to the ultrametric space $X$, and
$t$ (time) is a real parameter. For the investigation of this
equation we use the methods of ultrametric wavelet analysis, see the
Appendices 1 and 3 for the discussion.

We show that the above equation possesses solutions in the form of
the product of ultrametric wavelet and exponent of time
$$
v(x,t)=e^{\omega t}\psi_{Ij}(x).
$$

Moreover, we construct also more general solutions for the equation
above. We construct the unique solution of Cauchy problem with the
initial condition in the space $D_0(X)$ of mean zero test functions.
These solutions are built by means of the recurrent hierarchical
procedure. Procedures of this kind are related to the cascade models
of turbulence. Using this observation we will call the equation
above the quadratic cascade equation. Ultrametric ($p$--adic)
analysis was applied for the first time to description of cascade
models of turbulence in \cite{FischZelenov}.

The structure of the paper is as follows.

In Section 2 we construct solutions of Cauchy problem for the
quadratic cascade equation.

In Section 3 (the Appendix 1) we give the exposition of  ultrametric
analysis.

In Section 4 (the Appendix 2) we put the proof of the crucial Lemma
\ref{wavelet_interaction}.

In Section 5 (the Appendix 3) we discuss the cascade models of
turbulence.

\section{The quadratic cascade equation}

Consider the quadratic cascade equation
\begin{equation}\label{NSequ}
\frac{\partial}{\partial t}v(x,t)+\int\int F({\rm
sup}(x,\xi,\eta))v(\xi,t)\left(v(\eta,t)-v(x,t)\right)d\nu(\xi)
d\nu(\eta)+
$$
$$
+\int G({\rm sup}(x,\xi))(v(x,t)-v(\xi,t))d\nu(\xi)=0.
\end{equation}
Here the complex valued functions $F$ and $G$ on the tree ${\cal
T}(X)$ of balls in the (locally compact) ultrametric space $X$ with
the Borel measure $\nu$ should satisfy the condition of convergence
\begin{equation}\label{integrability}
\sum_{J>I}|F(J)|(\nu(J)-\nu(J,I))<\infty
\end{equation}
(analogously for $G$), where $\nu(J,I)$ is the measure of the
maximal subball in $J$ which contains the ball $I$.

The third term in the above equation contains the ultrametric
pseudodifferential operator, see the Appendix 1. The second term is
the integral operator with quadratic nonlinearity.

We have the following important lemma.

\begin{lemma}\label{wavelet_interaction}
\qquad{\sl Let $\phi$, $\psi$ be ultrametric wavelets. Denote $J$
(correspondingly $I$) the minimal ball which contains the support of
$\psi$ (correspondingly of $\phi$).

Let the integration kernel $F$ in (\ref{NSequ}) satisfies condition
(\ref{integrability}). Then the integral
\begin{equation}\label{I}
{\cal I}[\phi,\psi](x)=\int\int F({\rm
sup}(x,\xi,\eta))\phi(\xi)\left(\psi(\eta)-\psi(x)\right)d\nu(\xi)
d\nu(\eta)
\end{equation}
converges and  has the form:
$$
{\cal I}[\phi,\psi](x)=\psi(x)\phi(x)\Phi_{IJ},
$$
where the coefficient $\Phi_{IJ}$ can be non--zero only for the case
when $J<I$ (the ball $J$ is a strict subset of the ball $I$). In
this case
\begin{equation}\label{Phi_IJ}
\Phi_{IJ}=\nu^2(I,J)F(I)-\nu^2(J)F(J)
-\sum_{L:J<L<I}(\nu^2(L)-\nu^2(L,J))F(L).
\end{equation}

}

\end{lemma}

For the proof of the above lemma see Appendix 2.

\bigskip

\noindent{\bf Remark}\qquad If we fix in (\ref{I}) the order of
integration in the following way
$$
\int \left[\int F({\rm
sup}(x,\xi,\eta))\phi(\xi)\left(\psi(\eta)-\psi(x)\right)d\nu(\xi)\right]d\nu(\eta)
$$
then the integral (\ref{I}) will converge for any $\phi$, $\psi$
from $D_0(X)$ without the additional condition
(\ref{integrability}).

\bigskip

\noindent{\bf Example 1}\qquad  Using lemma
\ref{wavelet_interaction} one can easily  construct the particular
solution of the equation (\ref{NSequ}) in the form
$$
v(x,t)=e^{- \eta_{I} t}\psi(x)
$$
where $\psi$ is a wavelet (with the support in the ball $I$) and
$\eta_{I}$ is the corresponding eigenvalue for the
pseudodifferential operator in the third term of (\ref{NSequ}):
$$
\eta_{I}=G(I)\nu(I)+\sum_{J>I}G(J)(\nu(J)-\nu(J,I)).
$$
The introduced function will be a solution of (\ref{NSequ}) because
the integral ${\cal I}[\psi,\psi]$, where $\psi$ is a wavelet, is
identically zero.

\bigskip

\noindent{\bf Example 2}\qquad Let us find a solution of
(\ref{NSequ}) in the form
$$
v(x,t)=v_1(t)\psi(x)+v_2(t)\phi(x),
$$
where $\psi$ and $\phi$ are wavelets with the supports in the balls
$J$ and $I$ correspondingly.

If $I$ and $J$ have zero intersection or $I=J$ then ${\cal
I}[\phi,\psi]={\cal I}[\psi,\phi]=0$ and the corresponding solution
of (\ref{NSequ}) takes the form
$$
v(x,t)=v_1(0)e^{- \eta_{J} t}\psi(x)+v_2(0)e^{- \eta_{I} t}\phi(x).
$$

Let the balls $I$ and $J$ be comparable and different, for instance
let $J<I$. Then the corresponding solution of (\ref{NSequ}) takes
the form
$$
v(x,t) =v_1(0)e^{- \eta_{J}
t+\phi(x)\Phi_{IJ}v_2(0)\eta_I^{-1}\left( e^{- \eta_{I}
t}-1\right)}\psi(x)+v_2(0)e^{- \eta_{I} t}\phi(x).
$$
Here $\Phi_{IJ}$ is given by (\ref{Phi_IJ}) and we assume $\eta_I\ne
0$.

\bigskip

We have constructed the nontrivial solution of the nonlinear
equation (\ref{NSequ}). This solution contains a double exponent
(exponent of exponent) of time.

Let us consider Cauchy problem for equation (\ref{NSequ}) with
$v=v(x,t)\in D_0(X)\otimes C^1([0,\infty))$. The space
$D_0(X)\otimes C^1([0,\infty))$ is the inductive limit of spaces
$D_0({\cal S} )\otimes C^1([0,\infty))$. This means that any
function $v \in D_0(X)\otimes C^1([0,\infty))$ for any $t$ belongs
to some $D({\cal S})$ (where ${\cal S}$ does not depend on $t$) and
$v(x,t)$ is continuously differentiable with respect to $t$ for any
fixed $x$.

We say that $v\in D_0(X)\otimes C^1([0,\infty))$ is the solution of
Cauchy problem for (\ref{NSequ}) with the initial condition $v_0\in
D_0(X)$, if
$$
v(x,0)=v_0(x)
$$
and $v$ satisfies (\ref{NSequ}) for $t>0$.

The following theorem describes solutions of Cauchy problems for
(\ref{NSequ}) with initial conditions in $D_0(X)$.

\begin{theorem}\qquad
{\sl  Cauchy problem with the initial condition $v_0\in D_0(X)$ for
the integro--differential equation (\ref{NSequ}) which satisfies the
condition (\ref{integrability}) possesses the unique solution $v \in
D_0(X)\otimes C^1([0,\infty))$. }
\end{theorem}

\noindent{\it Proof}\qquad Any function $v \in D_0(X)\otimes
C^1([0,\infty))$ has the form of the finite linear combination of
wavelets
$$
v(x,t)=\sum_{Ij}v_{Ij}(t)\psi_{Ij}(x),
$$
where $v_{Ij}$ are in $C^1([0,\infty))$ and $I$ belong to some
${\cal S}\backslash {\cal S}_{\rm min}$, ${\cal S}\subset {\cal
T}(X)$. Substituting this into (\ref{NSequ}) we get
$$
\sum_{Ij}\psi_{Ij}(x)\left[\frac{d}{dt}v_{Ij}(t)+\eta_{I} v_{Ij}(t)
+ v_{Ij}(t)\sum_{Jj':J>I} v_{Jj'}(t)\psi_{Ji'}(x)\Phi_{JI}\right]=0.
$$
Since wavelets are linearly independent and wavelets of larger scale
are constants on the supports of wavelets of smaller scale, the
above equation is equivalent to the system of ordinary differential
equations
\begin{equation}\label{cascade}
\frac{d}{dt}v_{Ij}(t)=- v_{Ij}(t) \left[\eta_{I}  + \sum_{Jj':J>I}
v_{Jj'}(t)\psi_{Ji'}(x)\Phi_{JI} \right].
\end{equation}
Here the summation runs over the increasing sequence of balls which
are larger than $I$. This system is nonlinear (quadratic). Note
that, since $x\in I$ and the wavelet $\psi_{Jj}$ for $J>I$ is
constant on $I$, the coefficient $\psi_{Ji'}(x)$ does not depend on
the choice of $x\in I$.

Let us describe the recurrent procedure of construction of the
solution for the system (\ref{cascade}). The initial condition for
(\ref{NSequ}) as a function in $D_0(X)$ has the expansion over
wavelets
\begin{equation}\label{initial}
v(x,0)=\sum_{Ij}v_{Ij}(0)\psi_{Ij}(x).
\end{equation}
Let as choose a maximal $I$ for which the above initial condition
$v_{Ij}(0)$ is not equal to zero (such $I$ can be non unique). Since
for a maximal $I$ the corresponding equation in (\ref{cascade}) is
linear we obtain the corresponding exponential solution $v_{Ij}(t)$
of the Cauchy problem:
$$
v_{Ij}(t)=v_{Ij}(0)e^{-\eta_I t}.
$$

Then for maximal subballs $I'<I$ we substitute the obtained solution
in the corresponding equations in (\ref{cascade}) and obtain the
equations for $v_{I'j'}$. These equations will be {\it linear} and
will depend on the computed at the previous step function
$v_{Ij}(t)$.

Then we iterate this procedure and get solutions for all pairs
$(I,j)$:
\begin{equation}\label{recurrent}
v_{Ij}(t)=v_{Ij}(0)e^{-\eta_I t - \int_0^{t} \sum_{Jj':J>I}
v_{Jj'}(\tau)\psi_{Ji'}(x)\Phi_{JI} d\tau}.
\end{equation}
It is easy to see that if $v_{Ij}(0)=0$ then $v_{Ij}(t)=0$.
Therefore, since the initial condition is a finite linear
combination (\ref{initial}) of wavelets, the described recurrent
procedure will give all non zero $v_{Ij}(t)$ in a finite number of
steps.

The uniqueness of the solution follows from the observation that any
equation in (\ref{cascade}) is linear as equation for $v_{Ij}(t)$
and depends only on functions $v_{Jj'}(t)$ related to larger scales.
This finishes the proof of the theorem. $\square$

\bigskip

\noindent{\bf Remark}\qquad The formula (\ref{recurrent}) gives the
explicit solution of Cauchy problem for equation (\ref{NSequ}).
Therefore this equation is exactly solvable, and the solution can be
described with the help of the recurrent hierarchical procedure.

\bigskip

\noindent{\bf Remark}\qquad If the initial condition $v_0$ for
(\ref{NSequ}) belongs to some (finite dimensional) space $D_0({\cal
S})$, ${\cal S}\subset {\cal T}(X)$, then for any $t$ the solution
$v(x,t)$ of the corresponding Cauchy problem for (\ref{NSequ}) will
also belong to $D_0({\cal S})$. This observation is the example of
the mentioned in \cite{Doklady} general phenomenon of existence of
localized solutions for some integral equations in ultrametric
analysis.

\bigskip

\noindent{\bf Remark}\qquad The  equation (\ref{NSequ}) for $v$ in
$D_0(X)\otimes C^1([0,\infty))$ is equivalent to the system of
ordinary differential equations (\ref{cascade}). The system
(\ref{cascade}) gives the example of the so called cascade model.
Models of this kind are used for description of turbulence, see the
Appendix 3. The relation of nonlinear ultrametric integral equations
and cascade models of turbulence was mentioned for the first time in
\cite{FischZelenov}. For discussion of cascade models of turbulence
see \cite{Frick}.

The results of the present paper show that the cascade models are
related to integrable nonlinear ultrametric integral equations and
ultrametric wavelet analysis.

\section{Appendix 1: Ultrametric analysis}

In this Section we discuss some results on ultrametric analysis,
which can be found in  \cite{Izv}, \cite{ACHA}, \cite{MathSbornik}.

\begin{definition}\qquad{\sl
An ultrametric space is a set with the ultrametric $d(x,y)$ (where
$d(x,y)$ is called the distance between $x$ and $y$), i.e. a
function of two variables, satisfying the properties of positivity
and non degeneracy
$$
d(x,y)\ge 0,\qquad d(x,y)=0\quad \Longrightarrow\quad x=y;
$$
symmetricity
$$
d(x,y)=d(y,x);
$$
and the strong triangle inequality
$$
d(x,y)\le{\rm max }(d(x,z),d(y,z)),\qquad \forall x,y,z.
$$
}
\end{definition}

We say that an ultrametric space $X$ is regular, if this space
satisfies the following properties:

\medskip

1) The set of all the balls of nonzero diameter in $X$ is finite or
countable;

\medskip

2) For any decreasing (infinite) sequence of balls $\{I^{(k)}\}$,
$I^{(k)}\supset I^{(k+1)}$, the diameters of the balls tend to zero;

\medskip

3) Any ball of non--zero diameter is a finite union of maximal
subballs.

\bigskip

Ultrametric spaces are dual to directed trees. Below we describe
some part of the duality construction.

For a regular ultrametric space $X$ consider the set ${\cal T}(X)$,
which contains all the balls in $X$ of nonzero diameters, and the
balls of zero diameter which are maximal subbals in balls of nonzero
diameters. This set possesses a natural structure of a directed
tree. Two vertices $I$ and $J$ in ${\cal T}(X)$ are connected by an
edge if the corresponding balls are ordered by inclusion, say
$I\supset J$ (i.e. one of the balls contain the other), and there
are no intermediate balls between $I$ and $J$.

The partial order in ${\cal T}(X)$ is defined by inclusion of balls,
this partial order is a direction. We recall that a partially
ordered set is a directed set (and a partial order is a direction),
if for any pair of elements there exists the unique supremum with
respect to the partial order.

On the directed tree ${\cal T}(X)$ we have the natural increasing
positive function which puts into correspondence to any vertex the
diameter of the corresponding ball.

Assume now we have a directed tree ${\cal T}$ with the positive
increasing function $F$ on this tree. Then we define the ultrametric
on the set of vertices of the tree as follows: $d(I,J)=F({\rm
sup}(I,J))$ where ${\rm sup}(I,J)$ is the supremum of vertices $I$,
$J$ with respect to the direction.

Then we take completion of the set of vertices with respect to the
defined ultrametric and eliminate from the completion all the inner
points of the tree (a vertex of the tree is inner if it does not
belong to the border of the tree). We denote the obtained space
$X({\cal T})$, this space is ultrametric.

An ultrametric pseudodifferential operator is defined in the
following way. Consider a $\sigma$--additive Borel measure $\nu$
with countable or finite basis on a regular ultrametric space $X$.
Consider the pseudodifferential operator
$$
Tf(x)=\int T{({\rm sup}(x,y))}(f(x)-f(y))d\nu(y).
$$
Here $T{(I)}$ is some complex valued function on the tree ${\cal T}(
X)$. The supremum $ {\rm sup}(x,y)=I $ of the points $x,y\in X$ is
the minimal ball $I$ in $X$, containing both points.

\bigskip

Let us build a basis in the space $L^2(X,\nu)$ of complex valued
functions on a regular ultrametric space $X$ which are quadratically
integrable with respect to the measure $\nu$. We call this basis the
basis of ultrametric wavelets.

Denote by $V(I)$ the space of functions on $X$, generated by
characteristic functions of the maximal subballs in the ball $I$ of
nonzero diameter. Correspondingly, $V_0(I)$ is the subspace of
codimension 1 in $V(I)$ of functions with zero mean with respect to
the measure $\nu$. The spaces $V_0(I)$ for different $I$ are
orthogonal. We denote $p_I$ the number of maximal subballs in the
ball $I$. The dimension of $V_0(I)$ as the Euclidean space with the
scalar product as in $L^2(X,\nu)$  will be less or equal to $p_I-1$.

We introduce in the space $V_0(I)$ some orthonormal basis
$\{\psi_{Ij}\}$. If the measures of all maximal subballs in $I$ are
positive, the index $j$ can take values $1,\dots,p_I-1$. The next
theorem shows how to construct the orthonormal basis in
$L^2(X,\nu)$, taking the union of bases $\{\psi_{Ij}\}$ in the
spaces $V_0(I)$ over all non minimal $I\in {\cal T}(X)$
(equivalently, over all balls $I$ of non--zero diameters).

\begin{theorem}\label{basisX}\qquad
{\sl 1) Assume that the measure $\nu(X)$ of the regular ultrametric
space $X$ is infinite. Then the set of functions $\{\psi_{Ij}\}$,
where $I$ runs over all non minimal vertices of the tree ${\cal
T}(X)$ is an orthonormal basis in $L^2(X,\nu)$.

2) Let the measure $\nu(X)$ of the regular ultrametric space $X$ is
finite and is equal to $\nu(X)=A$. Then the set of functions
$\{\psi_{Ij}, A^{-\frac{1}{2}}\}$, where $I$ runs over all non
minimal vertices of the tree ${\cal T}(X)$ is an orthonormal basis
in $L^2(X,\nu)$.
 }
\end{theorem}

The basis introduced in the present theorem will be called the basis
of ultrametric wavelets.

The next theorem shows that the basis of ultrametric wavelets is the
basis of eigenvectors for ultrametric pseudodifferential operators.

\begin{theorem}\label{04}\qquad{\sl Let the following series converge absolutely:
\be\label{seriesconverge} \sum_{J>R} T{(J)} (\nu(J)-\nu(J,R))  \ee
for some ball $R$.

Then the ultrametric pseudodifferential operator
$$
Tf(x)=\int T{({\rm sup}(x,y))}(f(x)-f(y))d\nu(y)
$$
has a dense domain in $L^2(X,\nu)$ and ultrametric wavelets from
Theorem \ref{basisX} are eigenfunctions of $T$: \be\label{lemma2.1}
T\psi_{Ij}(x)=\lambda_I \psi_{Ij}(x) \ee with the eigenvalues:
\be\label{lemma4} \lambda_{I}=T{(I)} \nu(I)+\sum_{J>I} T{(J)}
(\nu(J)-\nu(J,I)). \ee Here $(J,I)$ is the maximal subball in $J$
which contains $I$.

}
\end{theorem}

A function $f$ on an (ultrametric) space $X$ is called locally
constant, if for any arbitrary point $x\in X$ there exists a
positive number $r$ (depending on $x$), such that the function $f$
is constant on the ball with the center in $x$ and the radius $r$:
$$
f(x)=f(y),\qquad \forall y: d(x,y)\le r.
$$

In particular, the characteristic function $\chi_I$ of a ball $I$ is
locally constant.

The space of test functions $D(X)$ on the ultrametric space $X$ is
defined as the space of locally constant functions with compact
support. Any test function in $D(X)$ is a (finite) linear
combination of characteristic functions of balls.

The space $D(X)$ is the inductive limit of the finite dimensional
spaces $D({\cal S})$, where ${\cal S}$ is a regular subtree in
${\cal T}(X)$ and $D({\cal S})$ is the linear span of characteristic
functions of balls from ${\cal S}$. Here the finite subtree ${\cal
S}\subset {\cal T}(X)$ is called regular, if  for any increasing
edge $IJ$, $I,J\in {\cal S}$, $I<J$ the subtree ${\cal S}$ contains
all edges $KJ$, $K<J$.

We denote $V_0({\cal S})\subset D({\cal S})$ the subspace of mean
zero functions in $D({\cal S})$. We denote $D_0({\cal S})$ the
factor space $V_0({\cal S})/V^{\perp}_0({\cal S})$ over the subspace
$V^{\perp}_0({\cal S})\subset V_0({\cal S})$ of functions which can
be not equal to zero only on some sets of zero measure $\nu$. The
space $D_0({\cal S})$ can be identified with the linear span of
ultrametric wavelets $\psi_{Ij}$, $I\in {\cal S}\backslash {\cal
S}_{\rm min}$, where ${\cal S}_{\rm min}$ is the set of the minimal
elements in ${\cal S}$.

We denote $D_0(X)$ the inductive limit of the finite dimensional
spaces $D_0({\cal S})$. The space $D_0(X)$ is isomorphic to the
linear span of ultrametric wavelets.

\section{Appendix 2: Proof of Lemma \ref{wavelet_interaction}}

Consider the integral
$$
{\cal I}[\phi,\psi](x)=\int\int F({\rm
sup}(x,\xi,\eta))\phi(\xi)\left(\psi(\eta)-\psi(x)\right)d\nu(\xi)
d\nu(\eta)
$$
for the case when $\phi$ and $\psi$ are ultrametric wavelets. The
condition (\ref{integrability}) guarantees the convergence of the
integral. Let us denote $I$ and $J$ the minimal balls which contain
the supports of $\phi$ and $\psi$ correspondingly.

Consider the following cases:

1) The balls $J$ and $I$ have empty intersection. Then the integral
over $\xi$ gives zero because $\phi$ is a mean zero function and the
integration kernel $F$ is constant on $I$.

2) Let $I$ be the strict subset of $J$. Then the integral over $\xi$
can be nonzero only for $x,\eta\in I$. In this case $\psi(\eta)$ and
$\psi(x)$ cancel and thus the integral is equal to zero.

3) Let $I=J$. Then the integral over $\xi$ can be nonzero only for
$x,\eta\in I$, and, moreover, ${\rm sup}(x,\eta)< I$. In this case
$\psi(\eta)$ and $\psi(x)$ cancel and thus the integral is equal to
zero.

4) Let $J$ be the strict subset of $I$. Consider the following
cases.

a) If $x\notin J$ then $\psi(x)=0$ and the integration kernel is
constant on $J$. Therefore the integral over $\eta$ is equal to
zero.

b) Let $x\in J<I$. The integral takes the form
$$
{\cal I}[\phi,\psi](x)=\int_{\xi\in I}\int_{\eta}F({\rm
sup}(x,\xi,\eta))\phi(\xi)\left(\psi(\eta)-\psi(x)\right)d\nu(\eta)
d\nu(\xi)=
$$
$$
=\int_{\xi\in I}\int_{\eta: {\rm sup}(x,\eta)< I}F({\rm
sup}(x,\xi,\eta))\phi(\xi)\left(\psi(\eta)-\psi(x)\right)d\nu(\eta)
d\nu(\xi)=
$$
$$
=\left[\int_{\xi:{\rm sup}(x,\xi)=I}+\int_{\xi:{\rm
sup}(x,\xi)<I}\right]\int_{\eta:  {\rm sup}(x,\eta)< I}F({\rm
sup}(x,\xi,\eta))\phi(\xi)\left(\psi(\eta)-\psi(x)\right)d\nu(\eta)
d\nu(\xi)=
$$
$$
=-\phi(x)\nu(I,J)F(I)\int_{\eta:  {\rm sup}(x,\eta)<
I}\left(\psi(\eta)-\psi(x)\right)d\nu(\eta)+
$$
$$
+\phi(x)\left[\int_{\xi:{\rm sup}(x,\xi)\le J}+\int_{\xi:J<{\rm
sup}(x,\xi)<I}\right]\int_{\eta:  {\rm sup}(x,\eta)< I}F({\rm
sup}(x,\xi,\eta))\left(\psi(\eta)-\psi(x)\right)d\nu(\eta)
d\nu(\xi)=
$$
$$
=\psi(x)\phi(x)\nu^2(I,J)F(I)+\phi(x)\nu(J)\int_{\eta: J\le {\rm
sup}(x,\eta)< I}F({\rm
sup}(x,\eta))\left(\psi(\eta)-\psi(x)\right)d\nu(\eta)-
$$
$$
-\psi(x)\phi(x)\int_{\xi:J<{\rm sup}(x,\xi)<I}\int_{\eta:  {\rm
sup}(x,\eta)< I}F({\rm sup}(x,\xi,\eta))d\nu(\eta) d\nu(\xi)=
$$
$$
=\psi(x)\phi(x)\biggl[\nu^2(I,J)F(I)-\nu^2(J)F(J) -\nu(J)
\sum_{L:J<L<I}F(L)(\nu(L)-\nu(L,J))-
$$
$$
-\sum_{L:J<L<I}(\nu(L)-\nu(L,J))\left(F(L)\nu(J)+F({\rm
sup}(L,K))\sum_{K:J<K<I}(\nu(K)-\nu(K,J))\right)\biggr]=
$$
$$
=\psi(x)\phi(x)\biggl[\nu^2(I,J)F(I)-\nu^2(J)F(J) -2\nu(J)
\sum_{L:J<L<I}F(L)(\nu(L)-\nu(L,J))-
$$
$$
-\sum_{L:J<L<I}(\nu(L)-\nu(L,J))\sum_{K:J<K<I}(\nu(K)-\nu(K,J))F({\rm
sup}(L,K))\biggr].
$$

Since
$$
\sum_{L:J<L<I}(\nu(L)-\nu(L,J))\sum_{K:J<K<I}(\nu(K)-\nu(K,J))F({\rm
sup}(L,K))=
$$
$$
=\sum_{L:J<L<I}(\nu(L)-\nu(L,J))^2F(L)+2\sum_{L:J<L<I}(\nu(L)-\nu(L,J))F(L)(\nu(L,J)-\nu(J)),
$$
we get for the integral
$$
{\cal
I}[\phi,\psi](x)=\psi(x)\phi(x)\biggl[\nu^2(I,J)F(I)-\nu^2(J)F(J) -
$$
$$
-\sum_{L:J<L<I}(\nu(L)-\nu(L,J))^2F(L)-2\sum_{L:J<L<I}(\nu(L)-\nu(L,J))F(L)\nu(L,J)\biggr]=
$$
$$
=\psi(x)\phi(x)\biggl[\nu^2(I,J)F(I)-\nu^2(J)F(J)
-\sum_{L:J<L<I}(\nu^2(L)-\nu^2(L,J))F(L)\biggr].
$$

This finishes the proof of the lemma. $\square$

\section{Appendix 3: Hierarchical models of turbulence}

In the present Section we discuss, following the book \cite{Frick},
the hierarchical (or cascade, or shell) models of turbulence. The
construction \cite{Frick} of cascade models is as follows.

1) The {\it hierarchical basis} in $L^2(\mathbb{R}^3)$, similar to
bases of wavelets, is constructed. The hierarchical basis is built
using translations and dilations of some finite set of mean zero
functions.

2) The hierarchical basis is applied to the investigation of the
Navier--Stokes equation
\begin{equation}\label{NS}
\rho \left(\frac{\partial \mathbf{v}}{\partial t} + \mathbf{v} \cdot
\nabla \mathbf{v}\right) = -\nabla p + \mu \nabla^2 \mathbf{v} +
\mathbf{f}.
\end{equation}
Here $\mathbf{v}$ is velocity, $\rho$ is density,
$p$ is pressure, $\mu$ is viscosity, and $\mathbf{f}$ is external
force.

This means that we substitute in the above equation the expansion of
velocity over the hierarchical basis, and investigate the obtained
system of non--linear ordinary differential equations for the
coefficients of the expansion:
\begin{equation}\label{exp_system}
\frac{\partial v_N(t)}{\partial
t}=\sum_{IJ}A_{NIJ}v_I(t)v_J(t)+\sum_{K}B_{NK}v_K(t)+C_N.
\end{equation}

3) We make the following important observation: in the obtained at
the previous step system of ordinary differential equations majority
of the coefficients will be small, and, moreover, will decay with
the distance from the main diagonal, i.e. $A_{NIJ}$ decays with
$|N-I|\to\infty$ or $|N-J|\to \infty$, $B_{NK}$ decays with
$|N-K|\to \infty$.

We define the cascade system of ordinary differential equations as
follows: we put in (\ref{exp_system}) the coefficients $A_{NIJ}$,
$B_{NK}$, corresponding to sufficiently large $|N-I|$, $|N-J|$,
$|N-K|$ equal to zero. We obtain the system of equations with
contributions corresponding to finite number of main diagonals of
the tensors $A_{NIJ}$, $B_{NK}$. In particular, any equation in the
obtained system will contain finite number of contributions. In this
sense the cascade model is constructed by cut--off of the system
(\ref{exp_system}).

The obtained in this way cascade systems of differential equations
were studied numerically \cite{Frick}. It was found that the
behavior of some cascade models is in good agreement with the
experimental results.

The hierarchical structure of the cascade models is well suited for
the description of the Richardson cascade and self--similarity of
turbulent flows \cite{Frick}. The indices of the hierarchical basis
in (\ref{exp_system}) correspond to eddies with different scales and
positions. Tensor $A_{NIJ}$ describes interaction of eddies, tensor
$B_{NK}$ describes dissipation of energy due to viscosity.

Let us discuss the relation of cascade models and ultrametric
analysis. The natural conjecture is that the cascade system of
ordinary differential equations is equivalent in some sense to the
single integral equation with integration over the ultrametric
argument. The first result of this kind was obtained in
\cite{FischZelenov}, where the $p$--adic integral equation which
describes the Richardson cascade was constructed.

In the present paper we introduce the approach related to
ultrametric wavelet analysis. We built the integro--differential
equation (\ref{NSequ}) with the following property: the system
(\ref{cascade}) obtained by expansion of $v(x,t)$ over the basis of
ultrametric wavelets is an example of the cascade system of
differential equations. In this approach the eddies correspond to
ultrametric wavelets, which itself are in correspondence with the
elements of the hierarchical basis of \cite{Frick}.  The second
(quadratic) term in (\ref{NSequ}) describes interaction of eddies,
the third term in (\ref{NSequ}) describes dissipation.  The cut--off
procedure used in the construction of the cascade model in our
approach takes the form of approximation of the Navier--Stokes
equation (\ref{NS}), in the regime of developed turbulence, by the
cascade equation (\ref{NSequ}). The approximation here is understood
as approximation of the system (\ref{exp_system}) by the system
(\ref{cascade}).

Actually in the present paper we have described by means of
ultrametric methods the simplest cascade model. We found that this
simplest cascade model is integrable. This shows the relation
between cascade models and theory of integrable nonlinear integral
ultrametric equations.

\bigskip\bigskip

\noindent{\bf Acknowledgments}\qquad The author would like to thank
E.I.Zelenov, I.V.Volovich, V.S.Vla\-di\-mi\-rov, A.Yu.Khrennikov and
S.Albeverio for fruitful discussions and valuable comments. The
author gratefully acknowledges being partially supported by the
grants DFG Project 436 RUS 113/809/0-1 and DFG Project 436 RUS
113/951, by the grants of The Russian Foundation for Basic Research
RFFI 05-01-04002-NNIO-a and RFFI 08-01-00727-a, by the grant of the
President of Russian Federation for the support of scientific
schools NSh 6705.2006.1 and by the Program of the Department of
Mathematics of Russian Academy of Science ''Modern problems of
theoretical mathematics''.

\end{document}